\documentclass[aps,pra,twocolumn,superscriptaddress,groupedaddress]{revtex4}
\usepackage{amssymb}
\usepackage{cancel}
\usepackage{color,graphicx}
\usepackage{amsmath}
\usepackage{amsbsy}
\usepackage{amsthm}
\usepackage{bbm}
\usepackage{epsfig}
\usepackage{lscape}
\usepackage{float}
\usepackage{graphicx}
\usepackage{subfigure}
\usepackage{dcolumn}
\usepackage{bbm}
\usepackage{color,epstopdf}
\usepackage{amscd}
\usepackage{amsfonts}  
\usepackage[]{amsmath}
\usepackage{amssymb}    
\usepackage{mathrsfs}
\usepackage{verbatim}
\usepackage[]{cases}
\usepackage{amsmath}
\usepackage{wasysym}
\usepackage[utf8]{inputenc}
\usepackage[T1]{fontenc}
\usepackage[usenames,dvipsnames,svgnames,table]{xcolor}

\newcommand{\braket}[1]{\langle {#1} \rangle}
\newcommand{\abs}[1]{\left| \,{#1} \right|}

\newcommand{\com}[2]{\left[ {#1}\, ,{#2} \right]}

\newcommand{\con}{\!\!-\!\!}
\newcommand{\1}{\mathbbm{1}}

\newcommand{\se}{\text{\tiny{$\mathcal{SE}$}}}
\newcommand{\s}{\text{\tiny{$\mathcal{S}$}}}
\newcommand{\e}{\text{\tiny{$\mathcal{E}$}}}

\newcommand{\nlprod}{\operatornamewithlimits{\overset{\rightharpoonup}{\prod}}} 

\newcommand{\av}[1]{\Tr_{\e}\!\!\left[{#1}\ro_{\e}\right]} 	

\newcommand\ro{\hat\rho}

\newcommand\Vo{\hat V}
\newcommand\Ao{\hat A}

\newcommand\phio{\hat\phi}

\newcommand\Mc{\mathcal{M}}

\newcommand\Tr{\mathrm{Tr}}

\begin{document}
\title{Recursive approach for non-Markovian time-convolutionless master equations}
\author{G. Gasbarri}
\email{giulio.gasbarri@ts.infn.it}
\affiliation{Department of Physics, University of Trieste, Strada costiera 11, 34151 Trieste, Italy}  
\author{L. Ferialdi}
\email{ferialdi@ts.infn.it}
\affiliation{Department of Physics, University of Trieste, Strada costiera 11, 34151 Trieste, Italy}
\affiliation{INFN, Sezione di Trieste, Strada costiera 11, 34151 Trieste, Italy}
\date{\today}
\begin{abstract}
We consider a general open system dynamics and we provide a recursive method to derive the associated non-Markovian master equation in a perturbative series. The approach relies on a momenta expansion of the open system evolution. Unlike previous perturbative approaches of this kind, the method presented in this paper provides a recursive definition of each perturbative term. Furthermore, we give an intuitive diagrammatic description of each term of the series,
which provides an useful analytical tool to build them and to derive their structure in terms of commutators and anticommutators. We eventually apply our formalism to the evolution of the observables of the reduced system, by showing how the method can be applied to the adjoint master equation, and by developing a diagrammatic description of the associated series.
\end{abstract}

\maketitle

\section{Introduction}
The investigation of open systems dynamics in quantum physics has constantly grown in recent years, pushed by the interest in developing new quantum technologies~\cite{qtech}.
Open quantum systems are generally described by non-Markovian dynamics, which account  for the memory of the interaction between the system and the environment surrounding it. 
Unlike approximated Markovian dynamics, that are always described by a master equation of the Lindblad type~\cite{Lin76}, non-Markovian dynamics in general cannot be recast in a unique explicit structure.
There is a vast literature on the formal investigation of non-Markovian dynamics~\cite{BreVac09,ChrKos10,Vac16,NZ1,deVAlo17,stoch}, in this paper we are interested in investigating those dynamics that are derived from underlying physical models, i.e. obtained by tracing out the degrees of freedom of a physical environment, provided that the initial state is factorized. Recently, a microscopic derivation has been provided for a specific class of non-Markovian maps~\cite{DioFer14}, namely those describing a system interacting with a bosonic bath that is completely characterized by its two-point correlation function. Notorious examples that fall in this category are, e.g., the non-Markovian Brownian motion~\cite{HPZ,Fer17b}, and the spin-boson model~\cite{Legetal87,sb}. Moreover, it has been shown that if one considers a system described by a bosonic quadratic Hamiltonian, it is possible to derive analytically the family of Gaussian, non-Markovian, completely positive master equations~\cite{Fer16}.
However, there are many physical systems that do not fall into the Gaussian ansatz. Interesting examples are state transfer in quantum information~\cite{Ceretal}, Brownian motion with non-linear coupling~\cite{HPZ2}, the donor-acceptor model~\cite{donor}, widely used in quantum biology; driven spin-chains~\cite{Pro} that cover a crucial role in condensed matter, and coupled cavities in cQED~\cite{Soletal}. 

General non-Markovian dynamics can be formally encoded in the Nakajima-Zwanzig master equation~\cite{BrePet02,NakZwa}, that displays an integral term accounting for memory effects. This class of master equation has been thoroughly investigated~\cite{NZ1,NZ2}, and only
recently, a characterization of physically admissible integro-differential master equations has been provided, based on a generalization of classical semi-Markov processes~\cite{Vac16}. Since integro-differential equations are hard to treat, a more handful tool to investigate open quantum systems are time convolutionless (TCL) master equations~\cite{BrePet02}. We underline that the solution of a TCL master equation always satisfies a Nakajima-Zwanzig master equation~\cite{ChrKos10}.
Closed expressions for TCL master equations have been obtained for few analytically solvable models~\cite{CalLeg,DLuc,HPZ} whose dynamics fall into the family of Gaussian non-Markovian maps~\cite{DioFer14,Fer16}. In order to derive TCL  master equations in more general frameworks, a number of perturbative approaches have been developed. Among these we mention the functional integral formalism~\cite{FeyVer}, the methods by Kubo and van Kampen~\cite{Kubo,vKa74, chat} (originally developed in the in the context linear stochastic differential equations), projection operator techniques~\cite{tclpof}, hierarchical equations of motion~\cite{heom}, effective modes~\cite{Cheetal14}, stochastic Liouville-Von Neumann~\cite{deV15}, and multiple-time correlation functions~\cite{deVAlo06} (for a review on the topic see~\cite{deVAlo17}).
These approaches allowed to improve the theoretical description of non-Markovian dynamics, but they all suffer of two drawbacks.
First, in order to obtain the series up to the $n$-th perturbative order, one has to apply repeatedly the whole formalism. This makes the derivation of higher order terms unwieldy.
Second, these methods do not make clear evidence of the mathematical structure of the perturbative series in terms of commutators and anticommutators.

In this paper, we tackle these issues by providing a perturbative technique that allows to derive the  master equation of a general open system in terms of a perturbative series, with the only assumption that the system and the bath are initially uncorrelated.
Unlike all perturbative approaches present in the literature, our method allows to characterize the structure of each expansion term through an explicit recursive formula. Such an iterative structure makes their derivation simpler.
We further provide an intuitive diagrammatic description of each term of the series,
which provides an useful analytical tool to build them and to infer 
their structure in terms of commutators and anticommutators.

We eventually apply our formalism to the evolution of the observables of the reduced system. We show how the method can be applied to the adjoint master equation, and we develop a diagrammatic description of the associated series.

\section{Non-markovian map and master equation}
We consider a system ($\mathcal{S}$) interacting with a generic environment ($\mathcal{E}$). 
The evolution of the open system density matrix $\ro_{\se}$ in the interaction picture is described by the Von-Neumann equation ($\hbar=1$)
\begin{align}\label{vn}
i \frac{\partial \ro_{\se}(t)}{\partial t} = \com{\hat{V}_{t}}{\ro_{\se}(t)}\,,
\end{align}
where $\Vo_t$ is a generic interaction Hamiltonian between the system and the environment. In order to simplify the calculations to come, we assume $\Vo_t$ to be factorized, i.e.
\begin{align}\label{eq:vni}
\hat{V}_{t}= \hat{A}_{t}\hat{\phi}_{t}\,,
\end{align}
where $\hat{A}_{t}$ and $\phio_{t}$ respectively are Hermitian system and environment operators. We however stress that the formalism presented holds for the most general $\hat{V}_{t}= \sum_i\hat{A}^i_{t}\hat{\phi}^i_{t}$.
It is convenient to introduce the left-right formalism denoting by a subscript L (R) the operators acting on $\ro$ from the left (right)~\cite{LR}.
The dynamical map $\Phi_{t}$ for the open system is obtained by formally solving Eq.~\eqref{vn}:
\begin{align}\label{eq:dmse}
\Phi_{t}\ro_{\se}= \mathcal{T}\left( e^{-i \int_{0}^{t}d\tau (\hat{V}_{\tau L}-\hat{V}_{\tau R})}\right) \ro_{\se}\,,
\end{align} 
where $\mathcal{T}( \cdot )$ denotes the time ordering operator. Since we are interested on the effective evolution of the system $\mathcal{S}$, we aim for the reduced dynamical map $\mathcal{M}_t$ that evolves the initial state of the system $\ro_\s$ to the state $\ro_\s(t)$ at time $t$ ($\ro_\s(t)\equiv\mathcal{M}_{t}\ro_{\s}$). This is obtained by tracing out the environmental degrees of freedom from $\Phi_t$. In order to do so, we assume that the open system initial state is factorized: $\ro_{\se}=\ro_{\s}\otimes\ro_{\e}$.
The dynamical map $\mathcal{M}_{t}(\cdot)$ is then given by
\begin{align}\label{eq:rds}
\mathcal{M}_{t}\ro_{\s} = \av{\mathcal{T}\left( e^{-i\int_0^t d\tau V^-_{\tau}} \right) \ro_{\s}\otimes}\,,
\end{align}
where $\Tr_{\e}$ denotes the partial trace over the environment, and we have defined
\begin{align}\label{eq:chi}
V^-_{\tau} \equiv \,(\hat{V}_{\tau L}-\hat{V}_{\tau R})
\end{align}
(for convenience superoperators are not denoted by a hat).
We observe that Eq.~\eqref{eq:rds} guarantees the complete positivity of the map, since it can be understood as the Kraus-Stinespring decomposition of $\mathcal{M}_{t}$~\cite{Sti55}.
When the environment is completely characterized by its two point correlation function $\Tr_{\e}[{\hat{\phi}^{i}_{t}\hat{\phi}^{j}_{s}\ro_{\e}}]$, the trace can be performed exactly and one obtains a closed Gaussian form for $\Mc_t$~\cite{DioFer14}.
If in addition the Hamiltonian is at most quadratic and the system operators obey linear Heisenberg equations of motion, one can exploit Wick's theorem and derive the exact master equation~\cite{Fer16}. Unluckily, in the general case we are considering, such techniques cannot be exploited, and one needs to tackle the problem from another perspective.

The formalism we use is based on an expansion over the map momenta, that is close to the cumulant expansion introduced by van Kampen~\cite{vKa74}. The advantage of our formalism is that it allows to construct recursively the master equation, while this is not possible with the van Kampen approach~\footnote{see Appendix D for explicit analysis}.   Since the derivation is rather involved, we refer the reader to the Appendixes for mathematical details.
We start by expanding Eq.~\eqref{eq:rds} in Dyson's series, obtaining
\begin{align}\label{Mt}
\mathcal{M}_{t}
&= 1+\sum_{n=1}^{\infty}(-i)^{n} \mu_{n,t}\,,
\end{align}
where $\mu_{n,t}$ are the integrated momenta 
\begin{align}\label{mu}
\mu_{n,t}\hat{\rho}_{\s}=\frac{1}{n!}\av{\mathcal{T}\left(\int_{0}^{t}d\tau V^-_{\tau} \right)^{n}\hat{\rho}_{\s}\otimes}\,.
\end{align}
We observe that the subscript $n$ of $\mu$ denotes the power of the superoperator $V^-$, i.e. the number of operators $\Ao$ and $\phio$ displayed by momentum. The subscript $t$ , denoting time dependence, will be dropped in the remainder of this paper for compactness of notation. Doing so, we implicitly assume that the momenta are evaluated at time $t$, unless otherwise explicitly stated.
In order to make this formula more transparent, we need to make explicit the dependence of $\mu_n$ over the system operators $\Ao$ and on the environment $n$-point correlation functions.
It is convenient to introduce a new pair of superoperators: $A^{+}\equiv \Ao_{L}+\Ao_{R}$  and $A^{-}\equiv\Ao_{L}-\Ao_{R}$ (analogous definitions hold for environment operators $\phio$). This notation is particularly convenient because one can associate to $A^+$ an anticommutator and to $A^-$ a commutator. It will then be immediately evident how these building blocks contribute to the structure of the master equation.
We consider the definition~\eqref{mu} of $\mu_n$ and we replace Eqs.~\eqref{eq:vni} and \eqref{eq:chi} in it.
The result in terms of $A^{\pm}$ and $\phi^{\pm}$ is
\begin{equation}
\mu_{n}\hat{\rho}_{\s}=\frac{1}{n!\,2^{n}}\av{\mathcal{T}\bigg( \int_{0}^{t} d\tau(A^+_{\tau}\phi^-_{\tau}+A^-_{\tau}\phi^+_{\tau})\bigg)^{n}\hat{\rho}_{\s}\otimes}\,.
\end{equation}
It is important to observe that a $+$ superoperator for the systems is always associated to a $-$ superoperator for the environment, and vice versa. We will shortly show that this \lq\lq sign conservation rule\rq\rq~covers a crucial role for trace preservation of the map.
We now exploit the binomial theorem, and we make explicit the time ordering simply by conditioning the time integrals. After some manipulation one finds that the result of this procedure is the mixing of the $\pm$ superoperators (see Appendix A): 
\begin{align}\label{mufin}
\mu_{n}= \int_{0}^{t}d\bar{\tau}_n\sum_{j=1}^{n}\sum_{\mathcal{P}_j}A^-_{\tau_{1}}\dots A^{k_n}_{\tau_{n}}\, D^{+\dots\bar{k}_n}_{\tau_{1}\dots\tau_{n}}\,,
\end{align}
where $\int_0^td\bar{\tau}_n=\prod_{i=1}^n\int_0^t d\tau_i$, and $\mathcal{P}_j$ denotes all the permutations of the indexes $k_i\in\{+,-\}$, with $k_1=-$ and such that there is a $j$ number of minus superoperators. 
We have also introduced the bath \lq\lq ordered correlation functions\rq\rq, defined by
\begin{align}\label{ordcorr}
D^{+\dots\bar{k}_n}_{\tau_{1}\dots\tau_{n}}\equiv\frac{1}{2^{n}}\av{\phi^+_{\tau_{1}}\theta_{\tau_1\tau_2}\phi^{\bar{k}_2}_{\tau_{2}}\dots\theta_{\tau_{n-1}\tau_n}\phi^{\bar{k}_n}_{\tau_{n}}}\, ,
\end{align}
where $\theta_{\tau_i\tau_j}$ is one for $\tau_i>\tau_j$, zero otherwise, and  provides the ordering both of the operators $\phio$ in $D$, of the operators $\Ao$ in Eq.~\eqref{mufin}, by conditioning the integrals limits.
The $2^{n}$ prefactor represents the number of permutations of the operators $\phio$  contained in $D$ (provided by commutators $\phi^{-}$ and anticommutators $\phi^{+}$).
Moreover, $\bar{k}_i\equiv- k_i$ guarantees the sign conservation rule.
We now exploit the cyclicity property of the trace, that implies $\av{\phi^{-}\hat{O}}=0$ for any operator $\hat{O}$. According to the sign conservation rule, the contributions where $A^+$ is the first superoperator on the left are suppressed. As a consequence, the first system superoperator on the left of $\mu_{n}$ is always $A^-$. This is an important feature because it guarantees that the map is trace preserving (indeed $\mathrm{Tr}_{\s}[A^{-}\hat{O}\,\ro_{\s}]=0$).

Equation~\eqref{mufin} shows that the momenta $\mu_{n}$ are composed by the sum of all the permutations of  the products of $n$ superoperators $A^{\pm}$, where the first term on the left is $A^{-}$, and the associated environment correlation function is obtained by the sign conservation rule. 
By replacing Eq.~\eqref{mufin} in Eq.~\eqref{Mt} we obtain the explicit expression for the perturbative series of the map $\mathcal{M}_{t}$.
 We observe that, unlike the Gaussian case~\cite{DioFer14}, one cannot sum the series and is left with the formal expression~\eqref{Mt}. However, if we consider a Gaussian bath, we can decompose higher order correlation functions in~\eqref{mufin} by means of the Isserlis' theorem~\cite{Iss18}, and recover known results.

The dependence of the momenta on the system operators and the environment correlation function, is not only important for the map structure, but plays also a relevant role for the derivation of the master equation.
We look for a time local master equation of the type
\begin{align}\label{eq:master}
\partial_{t}{\ro}_\s(t)=\mathbb{L}_{t}\ro_\s(t)\,,
\end{align}
where the generator $\mathbb{L}_{t}$ can be formally written as follows: 
\begin{align}\label{eq:generator}
\mathbb{L}_{t}=\partial_{t}\left({\mathcal{M}}_{t}\right)\mathcal{M}_{t}^{-1}.
\end{align}
Exploiting the identity  $(1+x)^{-1}= \sum_{n=0}^{\infty}(-)^{n}x^{n}$ (under the assumption that $|\!|\mathcal{M}_{t}-1|\!|<1$) one can invert Eq.~\eqref{Mt} obtaining
\begin{align}\label{eq.m1}
\mathcal{M}_{t}^{-1}=& \sum_{n=0}^{\infty}(-i)^{n}M_{n}\,,
\end{align} 
where the superoperators $M_n$ are recursively defined as follows:
\begin{align}\label{eq:L_q}
M_{n}&=-\sum_{k=1}^{n}\mu_{k}M_{n-k} 
\,, 
\end{align}
with $M_{0}=1$ (see Appendix B). One then sees that $M_n$ is the sum of $\mu_{n}$ plus  the  products of momenta with order lower than $n$. Accordingly, $M_n$ and $\mu_n$ contain the same number $n$ of operators, while they differ for how the bath operators are clustered by $\mathrm{Tr}_\mathcal{E}$ (and ordered by $\mathcal{T}$). Indeed, in $\mu_n$ the operators $\phio$ are grouped together under the trace in Eq.~\eqref{ordcorr}, while in $M_n$ one needs to consider all possible clusterings of $n$ elements. This fact can be seen by replacing Eq.~\eqref{mufin} in Eq.~\eqref{eq:L_q}. For example, for $n=2$ one finds
\begin{align}\label{}
\mu_2&= \int_{0}^{t}\!\!\!d\bar{\tau}_2\, A^{-}_{\tau_1}A^-_{\tau_2}\, D^{++}_{\tau_1\tau_2}+A^{-}_{\tau_1}A^+_{\tau_2}\, D^{+-}_{\tau_1\tau_2}\,,\nonumber\\
M_2&=\int_{0}^{t}\!\!\!d\bar{\tau}_2\, A^{-}_{\tau_1}A^-_{\tau_2}\, (D^+_{\tau_1}D^+_{\tau_2}-D^{++}_{\tau_1\tau_2}) - A^{-}_{\tau_1}A^+_{\tau_2}\, D^{+-}_{\tau_1\tau_2}\,.\nonumber
\end{align}
According to Eq.~\eqref{ordcorr}, while in $\mu_2$ both bath operators $\phi$ are clustered together (e.g. in $D^{++}_{\tau_1\tau_2}$), $M_2$ displays a term with a different clustering ($D^+_{\tau_1}D^+_{\tau_2}$).

Replacing Eqs.~\eqref{Mt} and \eqref{eq.m1} in Eq.~\eqref{eq:generator}, and after some calculations one can find (see Appendix C)
\begin{align}\label{gen}
\mathbb{L}_{t}
&=\sum_{n=1}^{\infty}(-i)^{n}L_{n}\,,
\end{align}
with 
\begin{align}\label{Lq}
L_{n}&= \dot{\mu}_{n}-\sum_{k=1}^{n-1}L_{n-k}{\mu}_{k}\,,
\end{align}
 and $L_{0}=0$,  where we denoted the derivative with respect to time $t$ with a dot.
Equation~\eqref{Lq} is the recursive law that allows to build iteratively of each term of the expansion~\eqref{gen}.
Each $L_n$ is the sum of all the possible combinations of $A^\pm$ (provided that the first on the left is always $A^-$), suitably ordered and clustered. These terms are associated to peculiar combinations of ordered correlation functions, whose construction is elegantly described by the recursion~\eqref{Lq}. The first two terms of the series~\eqref{gen} can be easily obtained starting from the definition of the momenta~\eqref{mufin}: 
\begin{align}
\label{L1}L_{1}&=A^-_{t}\,D^+_{t}\,, \\
\label{L2}L_{2}&=A^-_{t}\!\!\int_{0}^{t}\!\!\!d\tau_{1}\bigg[A^+_{\tau_{1}}D^{+-}_{t\,\tau_{1}}+A^-_{\tau_{1}}\Big(D^{++}_{t\,\tau_{1}}-D^{+}_{t}\,D^{+}_{\tau_{1}}\Big)\!\bigg].
 \end{align}
However, the structure of the third term is already quite complicated, and higher order terms are rather involved to compute. In order to ease the computation of the generic $L_n$, we provide here an intuitive diagrammatic description of how they can be built.

\section{Diagrammatics} We introduce the following notation: 
\begin{align}\label{symb}
\con\,\Circle\,\con\,&=A^{+}_{\tau}\phi^-_{\tau}\,,\qquad\con\,\CIRCLE\,\con\,= A^{-}_{\tau}\phi^+_{\tau}\,,
\end{align}
and we represent the trace over the environmental degrees of freedom as linking the circles in the following way:
\begin{align}\label{rulen}
\underbrace{\CIRCLE\con\RIGHTcircle\con\dots\con\RIGHTcircle}_{n} =\int_{0}^{t} d \bar{\tau}_n\, A^{-}_{\tau_{1}}\dots A^{\pm}_{\tau_{n}} D^{+\dots\mp}_{\tau_{1},\dots\tau_{n}}\,,
\end{align}
where $\RIGHTcircle$  denotes that in any position one can put either $\Circle$ or $\CIRCLE$, and $D$ is the bath ordered correlation function defined in Eq.~\eqref{ordcorr}. We call the left hand side of Eq.~\eqref{rulen} \lq\lq$n$-th order connected diagram\rq\rq, while a \lq\lq $n$-th order non-connected diagram\rq\rq~is obtained by removing at least one connection (line connecting circles) from the respective connected diagram. 
Note that when we trace only a single symbol~\eqref{symb}, this simply results in dropping the side lines, i.e. the first order diagram reads:
\begin{align}\label{rule1}
\CIRCLE=\int_{0}^{t} d \tau_1 A^{-}_{\tau_{1}}D^+_{\tau_1}\, .
\end{align}
We stress that the role of the bath ordered correlation functions $D$ is to link together the circles, clustering and ordering them in a specific way. Accordingly, one has that, e.g.
\begin{align}
\label{conn3} \CIRCLE\con\CIRCLE\con\CIRCLE\,= \int_{0}^{t}d\bar{\tau}_3\,A^-_{\tau_{1}}A^-_{\tau_{2}}A^-_{\tau_{3}}\,D^{+++}_{\tau_1\tau_2\tau_3}
\end{align}
differs from
\begin{align}
\label{sconn3}\CIRCLE\con\CIRCLE\,\CIRCLE\,=\int_{0}^{t}d\bar{\tau}_3\,A^-_{\tau_{1}}A^-_{\tau_{2}}A^-_{\tau_{3}}\,D^{++}_{\tau_1\tau_2} D^+_{\tau_3}
 \end{align}
for how the bath operators are clustered in $D$.
Moreover, we observe that with this notation, the fact that the first superoperator on the left is always $A^-$ is rephrased as follows: the diagrams whose first circle on the left is white are null, i.e.
\begin{align}\label{white1}
\Circle\con\RIGHTcircle\con\RIGHTcircle\con\dots\con\RIGHTcircle=0\,.
\end{align}
Having introduced the basic elements of our diagrammatics, we can move to its application. We start from the map $\mathcal{M}_t$~\eqref{Mt}, which is defined in terms of the momenta $\mu_n$ of Eq.~\eqref{mufin}.
With the diagrams introduced above, one finds that the momentum $\mu_n$ is the sum of all possible $n$-th order connected diagrams, i.e.
\begin{align}
\mu_n=\sum_{j=1}^n\sum_{\mathcal{P}_j}\underbrace{\CIRCLE\con\RIGHTcircle\con\dots\con\RIGHTcircle}_{n}\,,
\end{align}
where now $\mathcal{P}_j$ denotes all permutations of black and white circles, such that there is a $j$ number of black ones.

In order to build the term $L_n$ of the generator, we denote the derivative with respect to $t$ with a dot over a circle. With this notation, one finds that the diagrammatic version of Eqs.~\eqref{L1}-\eqref{L2} reads
\begin{align}
L_{1}&=\dot{\CIRCLE}\,,\\
L_{2}&=\dot{\CIRCLE}\con\CIRCLE -\dot{\CIRCLE}\,{\CIRCLE}+\dot{\CIRCLE}\con\Circle\,.
\end{align} 
The procedure to build Eq.~\eqref{Lq} for a generic $n$ with this diagramatics is the following (we show the case $n=3$ as explicit example):
\begin{enumerate}
\item Write the $n$-th order connected diagram composed by $n$ black circles and put a dot on the first circle.\\
\begin{align}\label{conn3dot}
\dot{\CIRCLE}\con\CIRCLE\con\CIRCLE 
\end{align}
\item Remove a number $p\leq n-1$ of connections from the previous connected diagram, in all possible ways. Multiply the diagrams obtained at each step by $(-1)^p$. Repeat for all $p$, until all $n-1$ connections are removed, i.e. until all black circles are disconnected.
\begin{align}
 \dot{\CIRCLE}\con\CIRCLE\con\CIRCLE -\left(\,\dot{\CIRCLE}\con\CIRCLE\,\CIRCLE+\dot{\CIRCLE}\,\CIRCLE\con\CIRCLE\,\right)+\dot{\CIRCLE}\,\CIRCLE\,\CIRCLE
\end{align}

\item Turn a number $p\leq n-1$ of black circles of the diagrams obtained so far into white, in all possible ways, and remembering the rule~\eqref{white1}. Repeat for all $p$, until all circles (but the first) are white.
\begin{align}\label{dia3}
L_{3}&= \dot{\CIRCLE}\con\CIRCLE\con\CIRCLE -\dot{\CIRCLE}\con\CIRCLE\,\CIRCLE-\dot{\CIRCLE}\,\CIRCLE\con\CIRCLE+\dot{\CIRCLE}\,\CIRCLE\,\CIRCLE\nonumber\\
&+\dot{\CIRCLE}\con\CIRCLE\con\Circle\ \hspace{1,42cm} -\dot{\CIRCLE}\,\CIRCLE\con \Circle\nonumber\\
&+\dot{\CIRCLE}\con\Circle\con\CIRCLE-\dot{\CIRCLE}\con\Circle\,\CIRCLE\nonumber\\
&+\dot{\CIRCLE}\con\Circle\con\Circle
\end{align}
\item Exploit Eqs.~\eqref{rulen}-\eqref{rule1} to translate the diagram obtained in operatorial form. Equation~\eqref{dia3} reads
\begin{align}\label{L3}
L_{3}=\!\int_{0}^{t}d\bar{\tau}_2&\bigg[ A^-_{t}A^-_{\tau_{1}}A^-_{\tau_{2}}\Big(
D^{+++}_{t\,\tau_1\tau_2}
 - D^{+}_{t}D^{++}_{\tau_1\tau_2}\nonumber\\
&\hspace{1.35cm}-D^{++}_{t\,\tau_1}D^{+}_{\tau_2}
+D^{+}_{t}D^{+}_{\tau_1}D^{+}_{\tau_2}\!\Big)\nonumber\\
 &+A^-_{t} A^-_{\tau_{1}}A^+_{\tau_{2}}\Big(
D^{++-}_{t\,\tau_1\tau_2}
-D^{+}_{t}D^{+-}_{\tau_1\tau_2}\Big)\nonumber\\
&+A^-_{t}A^+_{\tau_{1}}A^-_{\tau_{2}}\Big(
D^{+-+}_{t\,\tau_1\tau_2}
-D^{+-}_{t\,\tau_1}\,D^{+}_{\tau_2}\Big)
\nonumber\\
&+A^-_{t}A^+_{\tau_{1}}A^+_{\tau_{2}} 
D^{+--}_{t\,\tau_1\tau_2} \bigg]\,,
\end{align}
where 
 the first two lines correspond to the first line of Eq.~\eqref{dia3}.
\end{enumerate}
Equation~\eqref{L3} clearly provides insight on the mathematical structure of the master equation: $L_n$ is the sum of all possible combinations of commutators ($A^-$) and anticommutators ($A^+$) of operators $\Ao$, multiplied by suitable combinations of bath ordered correlation functions $D$, that encode the environment influence over the system.
If one considers a Gaussian bath, one can decompose any even ordered correlation function (odd ones are zero) in terms of the two point correlation function. Moreover, if the system Hamiltonian is bosonic and quadratic, one can exploit the operators algebra and reduce combinations of $n$ nested (anti-)commutators to double (anti-\!\!~)commutators, recovering the results of~\cite{Fer16}.

\section{Adjoint master equation} We derive the adjoint master equation for the system observables, that is useful tool to investigate the evolution of physical quantities.

We define the adjoint interaction picture as the picture where the statistical operator $\ro_{\se}$ evolves according to the free dynamics, and a generic operator $\hat{O}$ evolves with the interaction Hamiltonian, \textit{i.e.}
\begin{align}
\hat{O}_{t}&= \Phi_{t}^{*}\hat{O}, \nonumber\\
\ro_{\se}(t)&= e^{-i\left({H}^{-}_{\s}+{H}^{-}_{\e}\right)t}\ro_{\se}
\end{align} 
where ${H}^{-}_{\s}$ and ${H}^{-}_{\e}$ respectively are the generators of the free dynamics of the system and of the environment (defined like in Eq.~\eqref{eq:chi}), and $\Phi^{*}$ is the  adjoint dynamical map defined by
\begin{align}
\Phi_{t}^{*} \equiv\mathcal{T}\left( e^{i \int_{0}^{t}d\tau {V}_{\tau}^{-}}\right).
\end{align}
Since we are interested in the effective evolution of the system $\mathcal{S}$, we restrict our attention to operators of the type $\hat{O}=\hat{O}_{\s}\otimes \hat{\1}_{\e}$.
Under this assumption, we obtain the reduced dynamical map $\mathcal{M}_{t}^{*}$ by tracing out the environmental degrees of freedom from $\Phi_{t}^{*}$.
Under the further assumption that the initial state is factorized, the map $\mathcal{M}_{t}^{*}$ is  given by:
\begin{align}\label{eq:adjmap}
\mathcal{M}_{t}^{*}\hat{O}_{\s}= \Tr_{\e}\left[\ro_{\e}(t) \mathcal{T}\left( e^{i\int_0^t d\tau V^-_{\tau}} \right)\right]\hat{O}_{\s}.
\end{align}
Retracing the steps that we have done deriving the master equation \eqref{eq:master}, we expand Eq.~\eqref{eq:adjmap} in Dyson series, obtaining:
\begin{align}\label{eq:admap}
\mathcal{M}^{*}_{t}= 1+\sum_{n=1}^{\infty}i^{n}\tilde{\mu}_{n}\,,
\end{align}
where $\tilde{\mu}_{n}$ are the \lq\lq adjoint integrated momenta\rq\rq
\begin{align}\label{amufin}
\tilde{\mu}_{n}= \int_{0}^{t}d\bar{\tau}_{n}\sum_{j=1}^{n}\sum_{\mathcal{P}_{j}}{A}^{k_{1}}_{\tau_{1}}\dots {A}^{-}_{\tau_{n}}\tilde{D}_{\tau_{1}\dots\tau_{n}}^{\bar{k}_{1}\dots+}\,,
\end{align}
and the \lq\lq adjoint ordered correlation function\rq\rq~is defined by
\begin{align}
\tilde{D}_{\tau_{1}\dots\tau_{n}}^{\bar{k}_{1}\dots+}\equiv\,& \frac{1}{2^{n}}\Tr_{\e}\left[\ro_{\e}(t)\phi^{\bar{k}_{1}}_{\tau_{1}}\theta_{\tau_1\tau_2}\phi^{\bar{k}_2}_{\tau_{2}}\dots\theta_{\tau_{n-1}\tau_n}\phi^{+}_{\tau_{n}}\right]\,.
\end{align}
The adjoint momenta defined in Eq.~\eqref{amufin} differ from the momenta in Eq.~\eqref{mufin} only for the enviromental contribution $\tilde{D}$.
We now look for an adjoint master equation of the type
\begin{align}
\partial_{t}\hat{O}_{\s,t}= \mathbb{L}^{*}_{t}\hat{O}_{\s,t}.
\end{align}
The similarities between Eq.~\eqref{Mt} and Eq.~\eqref{eq:admap} allow to compute the generator $\mathbb{L}^{*}_{t}$ of the adjoint master equation, that is described by the series:
\begin{align}\label{agen}
\mathbb{L}^{*}_{t}
&=\sum_{n=1}^{\infty}i^{n}\tilde{L}_{n}\,,
\end{align}
with 
\begin{align}\label{aLq}
\tilde{L}_{n}&= \dot{\tilde{\mu}}_{n}-\sum_{k=1}^{n-1}\tilde{L}_{n-k}{\tilde{\mu}}_{k}
\end{align}
and $\tilde{L}_{0}=0$.
Because of the time dependence of the environmental state $\ro_{\e}$, we cannot directly implement the diagrammatic scheme developed for the master equation \eqref{eq:master}.
However, if we restrict our analysis to the steady states of the free evolution (${H}^{-}_{\e}\ro_{\e}=0$),
the environmental state $\ro_{\e}$ will  drop its time dependence.
 This allows to define a diagrammatic expression for the adjoint master equation by exploiting the  scheme previously developed, where the rule \eqref{rulen} is replaced by:
\begin{align}\label{rulend}
\underbrace{\RIGHTcircle\con\RIGHTcircle\con\dots\con\CIRCLE}_{n} =\int_{0}^{t} d \bar{\tau}_n\, A^{-}_{\tau_{1}}\dots A^{\pm}_{\tau_{n}} \tilde{D}^{+\dots\mp}_{\tau_{1},\dots\tau_{n}}\,,
\end{align}
and the rule \eqref{white1} is replaced by
\begin{align}\label{white2}
\RIGHTcircle\con\RIGHTcircle\con\RIGHTcircle\con\dots\con\Circle=0\,.
\end{align}

\section{Conclusions} 
 We have provided an iterative method that allows to derive in a perturbative series the non-Markovian master equation for the density matrix, and its adjoint for the observables, of a generic open quantum system.  The merit of our  formalism is that the expansion terms are defined recursively, making their derivation easier compared to previous perturbative techniques. We have further given a diagrammatic description of the expansion terms, that provides an intuitive analytical tool to build the perturbative series. Such a diagrammatics gives clear evidence of the mathematical structure of each term of the series, and explicitly shows that the environmental effects on the dynamics are encoded on the action of a series of commutators and anti-commutators of system operators, connected by the $n$-point environmental correlation functions. 


\section*{Acknowledgements} 
 The work of L.F. was supported by the TALENTS$^3$ Fellowship Programme, CUP code J26D15000050009, FP code 1532453001, managed by AREA Science Park through the European Social Fund.

\section*{Appendix A: Explicit derivation of the effective map.}
In this section we derive the explicit expression (9) for the ordered momenta $\mu_{n}$.
We consider Eq.~(8) and exploiting the binomial theorem we rewrite the momentum $\mu_{n}$ as follows:
\begin{widetext}
\begin{align}
\mu_{n}\,\hat{\rho}_{\s}
&=\frac{1}{2^{n}n!}\sum_{k=0}^{n}\frac{n!}{(n-k)!k!}\av{\mathcal{T} \left(\int_{0}^{t}d \tau A^+_{\tau}\phi^-_{\tau}\right)^{k}\left(\int_{0}^{t}d \tau A^-_{\tau}\phi^+_{\tau}\right)^{n-k}\hat{\rho}_{\s}\otimes}.
\end{align}
In order to make the time ordering explicit, we adopt the following strategy: we first resolve the time ordering for the couples $A^{+}\phi^{-}$ and $A^{-}\phi^{+}$ independently, by conditioning the integrals with unit step functions $\theta$:  
\begin{align}
\mu_{n}\,\hat{\rho}_{\s}&=\frac{1}{2^{n}n!}\sum_{k=0}^{n}\frac{n!}{(n-k)!k!}\av{\mathcal{T}\bigg[\mathcal{T}(\prod_{j=k+1}^{n}A^+_{\tau_{j}}\phi^{-}_{\tau_{k}})\mathcal{T}(\prod_{i=1}^{k}A^{-}_{\tau_{i}}\phi^{+}_{\tau_{i}})\bigg]\hat{\rho}_{\s}\otimes}\nonumber\\
&=\frac{1}{2^{n}}\sum_{k=0}^{n}\av{\mathcal{T}\nlprod_{j=1}^{n-k}\int_{0}^{t} d\tau_{j}(A^{+}_{\tau_{j}}\phi^{-}_{\tau_{j}})\theta_{\tau_{j-1},\tau_{j}}\nlprod_{i=1}^{k}\int_{0}^{t} d\tau_{i}(A^{-}_{\tau_{i}}\phi^{+}_{\tau_{i}})\theta_{\tau_{i-1},\tau_{i}}\,\hat{\rho}_{\s}\otimes}\,,
\end{align}
\end{widetext}
where $\tau_{0}=0$, and the arrow above the product denotes that the superoperators are ordered from the left to the right.
One can see that this partial time ordering removes the factorial terms in the equation and orders in two independent blocks the integrals associated to the two couples of operators. The second step of our derivation is to order globally the two \lq\lq pre-ordered\rq\rq~blocks. The result of this further ordering is the mixing of plus and minus superoperators in all the possible permutations, and the ordering of all integrals: 
\begin{align}
\mu_{n}\,\hat{\rho}_{\s}=& \frac{1}{2^{n}}\int_{0}^{t}d\bar{\tau}_n\sum_{j=1}^{n}\sum_{\mathcal{P}_j}A^{k_{1}}_{\tau_{1}}\dots A^{k_n}_{\tau_{n}}\\
&\hspace{0.5cm}\times \av{\phi^{\bar{k}_{1}}_{\tau_{1}}\theta_{\tau_1\tau_2}\phi^{\bar{k}_2}_{\tau_{2}}\dots\theta_{\tau_{n-1}\tau_n}\phi^{\bar{k}_n}_{\tau_{n}}}\,\hat{\rho}_{\s}\nonumber
\end{align}
where $d\bar{\tau}_n=\prod_{i=1}^n d\tau_i$, and $\mathcal{P}_j$ denotes all the permutations of the indexes $k_i\in\{+,-\}$ such that there is a $j$ number of minus superoperators and $\bar{k}_{i}=-k_{i}$.
Defining the environment ordered correlation function as
\begin{equation}\label{ordcorrapp}
D^{\bar{k}_{1}\dots\bar{k}_n}_{\tau_{1}\dots\tau_{n}}\equiv \frac{1}{2^{n}}\av{\phi^{\bar{k}_{1}}_{\tau_{1}}\theta_{\tau_1\tau_2}\phi^{\bar{k}_2}_{\tau_{2}}\dots\theta_{\tau_{n-1}\tau_n}\phi^{\bar{k}_n}_{\tau_{n}}}\,,
\end{equation}
we recover Eq.~(9):
\begin{align}
\mu_{n}= \int_{0}^{t}d\bar{\tau}_n\sum_{j=1}^{n}\sum_{\mathcal{P}_j}A^{k_{1}}_{\tau_{1}}\dots A^{k_n}_{\tau_{n}}\,D^{\bar{k}_{1}\dots\bar{k}_n}_{\tau_{1}\dots\tau_{n}}\,.
\end{align}

\section*{Appendix B: Recursive series for the inverse map $\mathcal{M}_{t}^{-1}$}

Aim of this section is to explicitly derive Eqs.~(13)-(14), i.e. to express $\mathcal{M}_{t}^{-1}$ as a power series of the interaction Hamiltonian $\Vo_{t}$ of Eq.~(2). 
We consider Eq.~(7) and we formally invert it, in such a way that $\mathcal{M}_{t}^{-1}$ (when it exists) can be written as:
\begin{align}
\mathcal{M}_{t}^{-1}= \left(1+\sum_{n=1}^{\infty}(-i)^{n}\mathcal{\mu}_{n}\right)^{-1}.
\end{align}
We define $\sum_{n=1}^{\infty}(-i)^{n}\mathcal{\mu}_{n}\equiv x$ and, assuming that $\abs{x}\le 1$, we exploit the identity $(1+x)^{-1}=\sum_{n=0}^{\infty}(-)^{n}x^{n}$ obtaining:
\begin{align}
\mathcal{M}_{t}^{-1}
=& \sum_{n=0}^{\infty}\left(-\sum_{k=1}^{\infty}(-i)^{k}\mu_{k} \right)^{n}.
\end{align}
This equation can be rearranged by making explicit the power $n$:
\begin{align}\label{invM}
\mathcal{M}_{t}^{-1} 
=&\sum_{n=0}^{\infty}(-)^{n}\prod_{j=1}^{n}\left(\sum_{k_{j}=1}^{\infty}(-i)^{k_{j}}\mu_{k_{j}}\right).
\end{align} 
From this expression it is clear that each term of the first series on the left is the product of series of momenta. Accordingly, such a series is not a power series of the interaction Hamiltonian $\Vo_{t}$ because each of its terms contains all powers of momenta (and hence of the interaction $\Vo_{t}$). We rearrange Eq.~\eqref{invM} by exploiting the Cauchy product of two series recursively (over the product of $n$ series). The result is:
\begin{align}\label{eq.m1app}
\mathcal{M}_{t}^{-1}=& \sum_{n=0}^{\infty}(-i)^{n}M_{n}
\end{align} 
with
\begin{align}\label{eq:Mq}
M_{n}=\sum_{q=0}^{n}(-)^{q}\sum_{k_{1}+\dots k_{q}=n}\mu_{k_{1}}\dots\mu_{k_q}\,.
\end{align}
Equation~\eqref{eq.m1app} is the correct series in power of the interaction we were looking for, as
$n$ denotes the power of the interaction Hamiltonian $\Vo_{t}$. The index $q$ in Eq.~\eqref{eq:Mq} instead denotes the number of partitions in which the operators are clustered (by the momenta $\mu_k$).
Since this expression for $M_{n}$ is rather involved, we rewrite the terms for $n\geq1$ as follows:
\begin{align}\label{eq.LqLk}
M_{n}=& \sum_{k=1}^{n}\mu_{k}\sum_{q=0}^{n-k}(-)^{q+1} \sum_{k_{1}+\dots +k_{q} =n-k}\mu_{k_{1}}\dots\mu_{k_{q}}
\end{align} 
One can easily check that the second series in this equation is simply Eq.~\eqref{eq:Mq} for $M_{n-k}$. This leads us to the recursive formula of Eq.~(14):
\begin{align}\label{eq:L_qapp}
M_{n}=\sum_{k=1}^{n}\mu_{k}(-M_{n-k})\qquad\text{with}\qquad M_{0}=1. 
\end{align}
\section*{Appendix C: Recursive formula for the time local generator}
In this section we provide the technical details for the derivation of Eq.~(16).
We start from Eq.~(12) and we substitute in it  Eqs.~(6) and (13), obtaining:
\begin{align}\label{eq:tcl}
\mathbb{L}_{t}
=&\sum_{n=1}^{\infty}\dot{\mu}_{n}\sum_{k=0}^{\infty}(-i)^{n+k}M_{k}
\end{align}
Similarly to the previous section, also this one is not a series in powers of the interaction Hamiltonian $\Vo_{t}$. In order to reach this goal, we exploit again the Cauchy product of two series, and we rearrange Eq.~\eqref{eq:tcl} as follows:
\begin{align}\label{eq:tclgenerator}
\mathbb{L}_{t}
&=\sum_{n=1}^{\infty}(-i)^{n}L_{n}
\end{align}
with
\begin{align}\label{eq:Lq}
L_{n}=\sum_{k=1}^{n}\dot{\mu}_{k}\,M_{n-k}\,,
\end{align}
and $M_{n}$ is determined by the recursive formula \eqref{eq:Mq}. 
Equation~\eqref{eq:Lq} is a useful expression of $L_n$ for performing a numerical analysis. 
However, it is more elegant to derive a recursive relation that involves only $L_{n}$ and the momenta $\mu$. We do so by adopting the same strategy exploited in the previous section.
We first replace Eq.~\eqref{eq:Mq} to obtain the following explicit expression:
\begin{align}\label{eq:lqe}
L_{n}&=\sum_{q=0}^{n}(-)^{q}\sum_{k_{0}+\dots k_{q}=n} \dot{\mu}_{k_{0}}\mu_{k_{1}}\dots\mu_{k_{q}}.
\end{align}
We then rearrange this sum as follows:
\begin{align}\label{eq:lqex}
L_{n}&=\dot{\mu}_{n}-\sum_{k=1}^{n-1}\left(\sum_{q=0}^{n-k}(-)^{q}\sum_{k_{0}+\dots k_{q}=n-k} \dot{\mu}_{k_{0}}\mu_{k_{1}}\dots\mu_{k_{q}}\right) \mu_{k}.
\end{align}
Comparing this expression for $L_{n}$ with Eq.~\eqref{eq:lqe}, one finds the desired recursive formula:
\begin{align}
L_{n}= \dot{\mu}_{n}-\sum_{k=1}^{n-1}L_{n-k}\mu_{k}\qquad\text{with}\qquad L_{0}=0. 
\end{align}

\section*{Appendix D: Van Kampen vs. recursive expansion}

In this section we compare the cost of our formalism to Van Kampen formalism. We build both of the perturbative series up to the fourth term ($L_4$) of the expansions.

Exploiting Van Kampen formalism, one finds that the generator $\mathbb{L}_{t}$ can be described by the series in Eq.~(15) with 
\begin{align}
L_{n}=\int_{0}^{t}d\tau_{1}\dots \int_{0}^{\tau_{n-1}}d\tau_{n} \braket{V^{-}_{t}V^{-}_{\tau_{1}}\dots V^{-}_{\tau_{n}}}_{oc}
\end{align}
where
\begin{align}\label{eq:cumul}
\braket{V^{-}_{t}V^{-}_{\tau_{1}}\dots V^{-}_{\tau_{n}}}_{oc}
\equiv&\sum (-1)^{q-1}\braket{V^{-}_{t}\dots V^{-}_{\tau_{i}}}\\
&\hspace{0.2cm}\times\braket{V^{-}_{\tau_{j}}\dots V^{-}_{\tau_{k}}}\braket{V^{-}_{\tau_{l}}\dots V^{-}_{\tau_{m}}}\braket{\dots}\nonumber\,.
\end{align}
While $\braket{\dots}= \av{\dots}$ and we have introduced the new notation $\braket{V^{-}_{t}V^{-}_{\tau_{1}}\dots V^{-}_{\tau_{n}}}_{oc}$, where the subscript $_{oc}$ stands for \lq\lq ordered cumulants\rq\rq. These are defined by the following rules:
write a string composed by the product of n $V^{-}_{\tau_{p_{i}}} $ super-operators in between the brackets. Partition the string into an arbitrary number of q substrings with $1<q<n$ by inserting angular brackets in between the $V^{-}$ of the original string. Multiply the resulting expression by the factor $(-1)^{q-1}$.
Concerning the time arguments $\tau_{j}$, they are organized as follows:
The first factor of the string is always $V^{-}_{t}$. The remaining $V_{\tau_{i}}^{-}$ display all the permutations of the time arguments $\tau_{1}, \tau_{2},\dots, \tau_{n-1}$ such that in each substring they are chronologically ordered. For example, in Eq.~\eqref{eq:cumul} one has $t\ge\dots t_{i}$, $t_{j}\ge \dots \ge t_{k}$, and $t_{l}\ge \dots \ge t_{m}$ (See [27] for further details). 

Following this prescription, we write the first four terms of the Van Kampen expansion:
 \begin{align}
L_{1}&=\braket{V^{-}_{t}}_{oc}=\braket{V^{-}_{t}}\,,\\
L_{2}&=\int_{0}^{t} d\tau_{1}\braket{V^{-}_{t}V^{-}_{\tau_{1}}}_{oc}=\int_{0}^{t} d\tau_{1}\Big(\braket{V^{-}_{t}V^{-}_{\tau_{1}}}-\braket{V^{-}_{t}}\braket{V^{-}_{\tau_{1}}}\Big)\, ,\nonumber
\end{align}
\begin{widetext}
 \begin{align}
L_{3}&=\int_{0}^{t}\!\!d\tau_{1}\!\int_{0}^{\tau_1}\!\!d\tau_{2}\braket{V^{-}_{t}V^{-}_{\tau_{1}}V^{-}_{\tau_{2}}}_{oc}\\
&=\!\!\int_{0}^{t}\!\!\!d\tau_{1}\!\!\int_{0}^{\tau_1}\!\!\!d\tau_{2}\Big(\!\braket{V^{-}_{t}V^{-}_{\tau_{1}}V^{-}_{\tau_{2}}}\!-\!\braket{V^{-}_{t}V^{-}_{\tau_{1}}}\braket{V^{-}_{\tau_{2}}}\!-\!\braket{V^{-}_{t}V^{-}_{\tau_{2}}}\braket{V^{-}_{\tau_{1}}}\!-\!\braket{V^{-}_{t}}\braket{V^{-}_{\tau_{1}}V^{-}_{\tau_{2}}}\!+\!\braket{V^{-}_{t}}\braket{V^{-}_{\tau_{1}}}\braket{V^{-}_{\tau_{2}}}\!+\!\braket{V^{-}_{t}}\braket{V^{-}_{\tau_{2}}}\braket{V^{-}_{\tau_{1}}}\!\Big)\nonumber
\end{align}
\begin{align}
L_{4}&=\int_{0}^{t}\!\!d\tau_{1}\!\int_{0}^{\tau_1}\!\!d\tau_{2}\!\int_{0}^{\tau_2}\!\!d\tau_{3}\braket{V^{-}_{t}V^{-}_{\tau_{1}}V^{-}_{\tau_{2}}V^{-}_{\tau_{3}}}_{oc}\\
&=
\int_{0}^{t}\!\!d\tau_{1}\!\int_{0}^{\tau_1}\!\!d\tau_{2}\!\int_{0}^{\tau_2}\!\!d\tau_{3}
\Big(\braket{V^{-}_{t}V^{-}_{\tau_{1}}V^{-}_{\tau_{2}}V^{-}_{\tau_{3}}}-\braket{V^{-}_{t}V^{-}_{\tau_{1}}V^{-}_{\tau_{2}}}\braket{V^{-}_{\tau_{3}}}-\braket{V^{-}_{t}V^{-}_{\tau_{1}}V^{-}_{\tau_{3}}}\braket{V^{-}_{\tau_{2}}}-\braket{V^{-}_{t}V^{-}_{\tau_{2}}V^{-}_{\tau_{3}}}\braket{V^{-}_{\tau_{1}}}\nonumber\\
&\hspace{2.8cm}-\braket{V^{-}_{t}V^{-}_{\tau_{1}}}\braket{V^{-}_{\tau_{2}}V^{-}_{\tau_{3}}}-\braket{V^{-}_{t}V^{-}_{\tau_{2}}}\braket{V^{-}_{\tau_{1}}V^{-}_{\tau_{3}}}-\braket{V^{-}_{t}V^{-}_{\tau_{3}}}\braket{V^{-}_{\tau_{1}}V^{-}_{\tau_{2}}}-\braket{V^{-}_{t}}\braket{V^{-}_{\tau_{1}}V^{-}_{\tau_{2}}V^{-}_{\tau_{3}}}\nonumber\\
&\hspace{2.8cm}+\braket{V^{-}_{t}V^{-}_{\tau_{1}}}\braket{V^{-}_{\tau_{2}}}\braket{V^{-}_{\tau_{3}}}+\braket{V^{-}_{t}V^{-}_{\tau_{1}}}\braket{V^{-}_{\tau_{3}}}\braket{V^{-}_{\tau_{2}}}+\braket{V^{-}_{t}V^{-}_{\tau_{2}}}\braket{V^{-}_{\tau_{1}}}\braket{V^{-}_{\tau_{3}}}+\braket{V^{-}_{t}V^{-}_{\tau_{2}}}\braket{V^{-}_{\tau_{3}}}\braket{V^{-}_{\tau_{1}}}\nonumber\\
&\hspace{2.8cm}+\braket{V^{-}_{t}V^{-}_{\tau_{3}}}\braket{V^{-}_{\tau_{1}}}\braket{V^{-}_{\tau_{2}}}+\braket{V^{-}_{t}V^{-}_{\tau_{3}}}\braket{V^{-}_{\tau_{2}}}\braket{V^{-}_{\tau_{1}}}-\braket{V^{-}_{t}}\braket{V^{-}_{\tau_{1}}}\braket{V^{-}_{\tau_{2}}}\braket{V^{-}_{\tau_{3}}}
-\braket{V^{-}_{t}}\braket{V^{-}_{\tau_{1}}}\braket{V^{-}_{\tau_{3}}}\braket{V^{-}_{\tau_{2}}}\nonumber\\
&\hspace{2.8cm}
-\braket{V^{-}_{t}}\braket{V^{-}_{\tau_{2}}}\braket{V^{-}_{\tau_{1}}}\braket{V^{-}_{\tau_{3}}}
-\braket{V^{-}_{t}}\braket{V^{-}_{\tau_{2}}}\braket{V^{-}_{\tau_{3}}}\braket{V^{-}_{\tau_{1}}}
-\braket{V^{-}_{t}}\braket{V^{-}_{\tau_{3}}}\braket{V^{-}_{\tau_{1}}}\braket{V^{-}_{\tau_{2}}}
-\braket{V^{-}_{t}}\braket{V^{-}_{\tau_{3}}}\braket{V^{-}_{\tau_{2}}}\braket{V^{-}_{\tau_{1}}}\Big).\nonumber
\end{align}
These explicit expressions show that the complexity of the expansions terms grows very quickly with the expansion order. Moreover, it is not possible to obtain a recursive law, forcing one to repeatedly apply the cumbersome prescription described above to derive any expansion terms.
In order to ease the comparison, we now write the same terms exploiting Eq.~(C.4):
\begin{align}
L_{1}&=\braket{V^{-}_{t}}\,,\hspace{3cm}L_{2}=\int_{0}^{t} d\tau_{1}\Big(\braket{V^{-}_{t}V^{-}_{\tau_{1}}}-\braket{V^{-}_{t}}\braket{V^{-}_{\tau_{1}}}\Big)\, ,\nonumber\\
L_{3}&=\int_{0}^{t}\!\!d\tau_{1}\!\int_{0}^{t}\!\!d\tau_{2}\Big(\braket{V^{-}_{t}V^{-}_{\tau_{1}}V^{-}_{\tau_{2}}}\theta_{\tau_{1}\tau_{2}}-\braket{V^{-}_{t}V^{-}_{\tau_{1}}}\braket{V^{-}_{\tau_{2}}}-\braket{V^{-}_{t}}\braket{V^{-}_{\tau_{1}}V^{-}_{\tau_{2}}}\theta_{\tau_{1}\tau_{2}}+\braket{V^{-}_{t}}\braket{V^{-}_{\tau_{1}}}\braket{V^{-}_{\tau_{2}}}\Big)\nonumber\\
L_{4}&= \int_{0}^{t}\!\!d\tau_{1}\!\int_{0}^{t}\!\!d\tau_{2}\!\int_{0}^{t}\!\!d\tau_{3}\,\Big(\braket{V^{-}_{t}V^{-}_{\tau_{1}}V^{-}_{\tau_{2}}V^{-}_{\tau_{3}}}\theta_{\tau_{1}\tau_{2}\tau_{3}\tau_{4}}-
\braket{V^{-}_{t}V^{-}_{\tau_{1}}V^{-}_{\tau_{2}}}\braket{V^{-}_{\tau_{3}}}\theta_{\tau_{1}\tau_{2}}
-\braket{V^{-}_{t}V^{-}_{\tau_{1}}}\braket{V^{-}_{\tau_{2}}V^{-}_{\tau_{3}}}\theta_{\tau_{1}\tau_{2}}\theta_{\tau_{3}\tau_{4}}\nonumber\\
&\hspace{3cm}-\braket{V^{-}_{t}}\braket{V^{-}_{\tau_{1}}V^{-}_{\tau_{2}}V^{-}_{\tau_{3}}}\theta_{\tau_{1}\tau_{2}\tau_{3}}+\braket{V^{-}_{t}V^{-}_{\tau_{1}}}\braket{V^{-}_{\tau_{2}}}\braket{V^{-}_{\tau_{3}}}+\braket{V^{-}_{t}}\braket{V^{-}_{\tau_{1}}V^{-}_{\tau_{2}}}\braket{V^{-}_{\tau_{3}}}\theta_{\tau_{1}\tau_{2}}\nonumber\\
&\hspace{3.cm}+\braket{V^{-}_{t}}\braket{V^{-}_{\tau_{1}}}\braket{V^{-}_{\tau_{2}}V^{-}_{\tau_{3}}}\theta_{\tau_{2}\tau_{3}}-\braket{V^{-}_{t}}\braket{V^{-}_{\tau_{3}}}\braket{V^{-}_{\tau_{2}}}\braket{V^{-}_{\tau_{1}}}\Big)\,,
\end{align}
\end{widetext}
where $\theta_{\tau_1\dots\tau_n}$ is one for $\tau_1>\dots>\tau_n$, zero otherwise. Here we has used the following expression for the momenta:
\begin{align}
\mu_{n}=\int_{0}^{t}d\bar{\tau}_{n} \braket{V^{-}_{t}V^{-}_{\tau_{1}}\dots V^{-}_{\tau_{n}}}\,\theta_{\tau_1\dots\tau_n}\,,
\end{align}
which is equivalent to Eq.~(9) of the main text, as one can easily check by resolving the time ordering in Eq.~(7).
By comparing Eqs.~(D.3)-(D.4) it is evident that when the order of the expansion grows, our method requires the computation of a lower number of terms: 4 against 6 for $L_3$, and 8 against 20 for $L_4$. 
What is even more important is that the expressions above allow for a recursive writing (see Eq.~(16)):
\begin{align}
L_{1}&=\dot{\mu}_{1}\nonumber\\
L_{2}&=\dot{\mu}_{2}+L_{1}\mu_{1}\nonumber\\
L_{3}&=\dot{\mu}_{3}-L_{2}\mu_{1}-L_{1}\mu_{2}\nonumber\\
L_{4}&=\dot{\mu}_{4}-L_{3}\mu_{1}-L_{2}\mu_{2}-L_{1}\mu_{3}
\end{align}
This recursion reduces even further the number of terms that need to be computed at each order, and allows for a diagramatic description that eases their construction.

\end{document}